\documentstyle[12pt,amssymb]{article}

\def\slashchar#1{\setbox0=\hbox{$#1$}           
   \dimen0=\wd0                                 
   \setbox1=\hbox{/} \dimen1=\wd1               
   \ifdim\dimen0>\dimen1                        
      \rlap{\hbox to \dimen0{\hfil/\hfil}}      
      #1                                        
   \else                                        
      \rlap{\hbox to \dimen1{\hfil$#1$\hfil}}   
      /                                         
   \fi}                                         %
\def\half{\textstyle{1\over 2}}
\def\nn{\nonumber}

\begin{document}

\begin{titlepage}

{\hspace*{\fill} UM-TH-95-09 \\
 \hspace*{\fill} hep-ph/9504420 \\
 \hspace*{\fill} April 1995 \\}

\bigskip\bigskip

\begin{center}
{\Large\bf Chaotic inflation and a radiatively generated intermediate scale
in the supersymmetric standard model\\}
\end{center}

\bigskip

\begin{center}
{\large Tony Gherghetta\footnote{tgher@umich.edu} \, and \,
Gordon L. Kane\footnote{gkane@umich.edu} \\}
\medskip\medskip
{\it Randall Laboratory of Physics,\\University of Michigan, \\Ann Arbor, MI
48109-1120\\}
\end{center}

\bigskip

\begin{abstract}
\baselineskip=16pt

We consider a phenomenological extension of the minimal supersymmetric
standard model which incorporates chaotic inflation and a radiatively
generated intermediate mass scale. Initially a period of chaotic inflation
is driven by a quartic potential associated with the right-handed electron
sneutrino. Supersymmetry relates the quartic coupling of the inflationary
potential to the electron Majorana neutrino Yukawa coupling, $h_1$.
The microwave background temperature anisotropy determines this coupling to
be $h_1\simeq 10^{-7}$, which is similar in magnitude to the electron Dirac
Yukawa coupling. A U$(1)$ Peccei-Quinn (PQ) symmetry is broken by radiative
corrections at an intermediate scale $\simeq 10^{12}$GeV when the universe
cools to a temperature $T\lesssim 10^3$GeV. This leads to an invisible
axion, a weak scale $\mu$-term and an electron Majorana neutrino mass
$M_{N_1}\simeq 10^5$GeV. A second inflationary period can also occur via a
flat-direction field. In this case the universe can be reheated to a
temperature $T_{RH}\simeq 10^6$GeV, without restoring PQ symmetry.
Baryogenesis will then occur via out-of-equilibrium neutrino decay.

\end{abstract}

\end{titlepage}

\baselineskip=18pt

\section{Introduction}

While the minimal supersymmetric standard model (MSSM) provides the most
promising extension of the successful standard model, it does not yet
encompass important ideas that would be expected of the complete low
energy effective theory. These include neutrino masses, baryogenesis,
inflation and the strong CP problem. Recently there have been a number of
interesting proposals which partly address these shortcomings. In particular
several papers by Murayama, Yanagida and collaborators have made significant
contributions \cite{msyy}; see also ref. \cite{shafi}. However a more
complete phenomenological model is lacking at present.

In this work we construct a phenomenological extension of the MSSM which
can successfully incorporate inflation, neutrino masses, baryogenesis and
axions. We build on the work of Murayama et al \cite{msyy,msy}, modifying
their approach in a way specified below. The resulting Lagrangian can
describe all of the usual supersymmetry phenomenology, including cold dark
matter, LEP data and BR($b\rightarrow s\gamma$). In addition it includes
neutrino masses via a see-saw mechanism (which provide the hot dark matter),
induces inflation with a sneutrino inflaton, incorporates baryogenesis and
accommodates the axion. Interestingly the parameters relevant to each of
these ideas are closely interrelated in our model. Furthermore, the
fine-tuning in the inflationary potential is no more worse than that of the
electron Yukawa coupling in the MSSM. We will postpone giving any detailed
calculations here, and instead outline how such an encompassing Lagrangian
can be constructed.

In order to incorporate chaotic inflation one needs a scalar field in the
theory to have an initial value much greater than $M_{Planck}$ in the early
universe. An unnatural fine-tuning required for gauge non-singlet fields
along D-term flat directions restricts the inflaton to be a gauge singlet
field such as a sneutrino. This idea of identifying the right-handed
sneutrino as the inflaton is due to Murayama et al \cite{msyy}, where it was
noted that the addition of the superpotential term $W=\half M {\hat N_i^c}
{\hat N_i^c}$ with a common Majorana mass $M\simeq 10^{13}$GeV coincides
with a successful implementation of chaotic inflation using a quadratic
scalar potential. However to solve the strong CP problem one expects the
Lagrangian in the early universe to be Peccei-Quinn (PQ) invariant. A
PQ-invariant Majorana term for the right-handed neutrinos can be written by
introducing a singlet superfield $\hat P$ with superpotential $W=\half h_i
{\hat N_i^c} {\hat N_i^c} {\hat P}$. This means that in the early universe
the inflationary potential will be quartic with a coupling $h_1^2$.
Anisotropic temperature fluctuations, $\delta T/T \simeq 10^{-5}$ in the
present universe then determine the Majorana Yukawa coupling $h_1\simeq
10^{-7}$ \cite{salopek}.

Neutrino masses via a see-saw mechanism will be generated when the scalar
component ${\tilde P}$ of the superfield $\hat P$ receives a vacuum
expectation value at an intermediate scale. Intermediate scale breaking in
the supersymmetric standard model was previously considered by Murayama,
Suzuki and Yanagida \cite{msy}. Radiative corrections from right-handed
neutrino loops break U$(1)_{PQ}$ by driving the squared mass of ${\tilde P}$
negative at an intermediate scale. This is similar to the normal radiative
electroweak symmetry breaking induced in the MSSM by the large top Yukawa
coupling. It turns out that a second singlet superfield ${\hat P}^\prime$ is
also needed to ensure an invisible axion. The PQ symmetry can only be broken
after inflation ends because during the inflationary epoch the inflaton
induces an effective ${\tilde P}$ mass, which dominates the radiative
corrections from neutrino loops. As the inflaton undergoes coherent
oscillations about its minimum, the oscillation amplitude falls off as
$R^{-1}$ ($R$ is the scale factor of the universe) for a quartic potential.
However as the universe is reheated, finite temperature corrections induce
a local minimum at $\langle{\tilde P}\rangle=0$ and the field ${\tilde P}$
can remain trapped there until $T\lesssim 10^3$GeV.

If a second period of inflation were to commence when the sneutrino is
oscillating to zero, the universe would then be supercooled below
$T\simeq10^3$GeV. The potential barrier at $\langle{\tilde P}\rangle =0$
would disappear and $\tilde P$ would drop to the true vacuum at $\langle{
\tilde P}\rangle\simeq 10^{12}$GeV. This second inflationary epoch can be
caused by a scalar field with amplitude ${\cal O}(M_{Pl})$. Typically these
scalar fields only have non-renormalisable inflaton couplings and are
associated with a flat direction of the supersymmetric theory. Their
amplitudes can be driven to values ${\cal O}(M_{Pl})$ during the first
inflationary epoch \cite{drt}. Thus it is likely that after the first
inflation period is over there exists some flat-direction field, $\eta$ with
an amplitude ${\cal O}(M_{Pl})$ which starts the second inflationary epoch.
In contrast to the initial period of chaotic inflation where $V(\phi)
\lesssim M_{Pl}^4$, the potential along the flat direction is $V\simeq
m_W^2 M_{Pl}^2 \simeq (10^{11} {\rm GeV})^4$, where $m_W\simeq
{\cal O}$(TeV). This has been referred to as `intermediate scale inflation'
\cite{banks} and conveniently coincides with the U$(1)_{PQ}$ symmetry
breaking.

When the second inflationary epoch ends, the universe is reheated to a
temperature $T_{RH}\simeq 10^6$GeV which is low enough to prevent restoring
PQ symmetry (at the local minimum $\langle{\tilde P}\rangle=0$). This
reheat temperature is high enough for baryogenesis to occur via the
out-of-equilibrium decay of the light electron Majorana neutrino ($N_1$).
The initial chaotic inflationary epoch with the right-handed electron
sneutrino inflaton solves the flatness and horizon problems and generates
the required density perturbations. In order not to wipe out the density
perturbations from the original inflationary epoch, we require that the
number of e-foldings, $N$ during the second period of inflation satisfy
$N\lesssim 30$ \cite{rt}. Note, however that the axion strings resulting
from the spontaneous symmetry breakdown will not be completely diluted
during the second inflationary epoch.

The main points of our model which differ from previous attempts are as
follows. Initially chaotic inflation occurs with a quartic potential
associated with the right-handed electron sneutrino. COBE data on the
temperature anisotropy then determine the electron Majorana Yukawa coupling
to be $h_1\simeq 10^{-7}$ which is no less fine-tuned than the electron
Dirac Yukawa coupling. When the universe is reheated, finite temperature
corrections induce a local minimum at $\langle{\tilde P}\rangle=0$ which
persists until $T\simeq 10^3$GeV. If instead a second inflationary epoch
occurs at an intermediate scale (via a flat-direction field $\eta$), the
universe will be supercooled below $T\simeq 10^3$GeV as $\langle
{\tilde N}_1^c\rangle\rightarrow 0$. Soft breaking terms then dominate and
radiatively generate an intermediate mass scale $\simeq 10^{12}$GeV. The
mass of the electron Majorana neutrino will typically be $M_{N_1}\simeq
10^5$GeV (rather than the more common value $M_{N_1}\simeq 10^{11}$GeV).
When the flat-direction field decays it can reheat the universe to a
temperature $T_{RH}\simeq 10^6$ GeV. All supersymmetry breaking effects are
parameterised by soft terms in the scalar potential and we do not consider
any effects that might arise from supergravity or string theory. The details
of our scenario are presented below.

\section{Chaotic inflation in the supersymmetric standard model}

Consider a PQ invariant extension of the MSSM which provides the framework
for our model of inflation. This extension was first considered by Murayama,
Suzuki and Yanagida \cite{msy}. If a right-handed neutrino field
${\hat N}^c$ is introduced into the MSSM, the possible Yukawa couplings in
the superpotential are
\begin{equation}
	W[\Phi]=  h_U^{ij} {\hat u}^c_i {\hat Q}_j {\hat H}_u
	+ h_D^{ij} {\hat d}^c_i {\hat Q}_j {\hat H}_d
	+ h_E^{ij} {\hat e}^c_i {\hat L}_j {\hat H}_d
	+ h_N^{ij} {\hat N}^c_i {\hat L}_j {\hat H}_u
\label{sp}
\end{equation}
where $\hat Q$, $\hat L$ and ${\hat H}_{u,d}$ are SU(2) doublet chiral
superfields and ${\hat u}^c$,${\hat d}^c$,${\hat e}^c$ and ${\hat N}^c$ are
SU(2) singlet chiral superfields. The labels $i,j$ are generation indices
and all group indices have been suppressed. Notice that to generate a
Majorana mass term for the right-handed neutrino only requires coupling
${\hat N}^c$ to a singlet superfield $\hat P$. However, with just the
superfield $\hat P$, the PQ symmetry is broken at the electroweak scale and
gives rise to a standard visible axion which has been ruled out
experimentally. This problem is avoided by introducing a second singlet
field, ${\hat P}^\prime$ which causes the PQ symmetry to be broken at an
intermediate scale and leads to an invisible axion \cite{msy}. Thus the most
general PQ invariant superpotential with an intermediate breaking scale is
given by
\begin{equation}
	W^\prime[\Phi]={\half} h_M^{ij} {\hat N}^c_i {\hat N}^c_j {\hat P}
	+{f\over M_{Pl}} {\hat P}^3 {\hat P}^\prime
	+{g\over M_{Pl}} {\hat P} {\hat P}^\prime {\hat H}_u {\hat H}_d ,
\label{pqsp}
\end{equation}
where $M_{Pl}$ is the Planck mass and the PQ charge assignments are $+1/2$
for $\hat Q$, $\hat L$, ${\hat u}^c$, ${\hat d}^c$, ${\hat e}^c$,
${\hat N}^c$, $-1$ for $\hat P$, ${\hat H}_{u,d}$ and $+3$ for
${\hat P}^\prime$. The total superpotential of our phenomenological model is
$W+W^\prime$. Notice that by introducing ${\hat P}^\prime$ one naturally
generates a coupling to the Higgs superfields, which ultimately becomes the
$\mu$-term of the MSSM.

For chaotic inflation to occur one needs an inflationary potential $V(\phi)
\lesssim M_{Pl}^4$ and a scalar field with an initial value $\phi(0) \gg
M_{Pl}$ \cite{linde}. The scalar potential resulting from $W+W^\prime$
restricts the amplitude of any gauge non-singlet scalar fields to be
${\cal O}(M_{Pl})$ because of unnatural fine-tunings along D-term flat
directions \cite{msyy}.\footnote{Also a flat inflationary potential is much
more difficult to achieve with gauge couplings.}
This leaves only the scalar components ${\tilde N}_i^c$, ${\tilde P}$ and
${\tilde P}^\prime$ of the singlet superfields ${\hat N}_i^c$,$\hat P$ and
${\hat P}^\prime$ as possible candidates for the inflaton. The scalar
potential arising from the superpotential $W^\prime$ is given by
\begin{eqnarray}
	V(\phi)&=&\left|{\half} h_i {\tilde N}_i^c{\tilde N}_i^c+3{f\over
	M_{Pl}}{\tilde P}^2{\tilde P}^\prime+{g\over M_{Pl}} H_uH_d
	{\tilde P}^\prime\right|^2+\left|{\tilde P}\right|^2 \left|{f\over
	M_{Pl}}	{\tilde P}^2+{g\over M_{Pl}}H_u H_d\right|^2 \nn \\
	&+& h_i^2\left|{\tilde N}_i^c\right|^2 \left|{\tilde P}\right|^2
	+{g^2\over M_{Pl}^2}\left|{\tilde P}\right|^2 \left|{\tilde P}^
	\prime\right|^2 (H_u^\dagger H_u+H_d^\dagger H_d)
\label{vphi}
\end{eqnarray}
where we have assumed for simplicity that the Majorana Yukawa couplings
are real and diagonal, $h_M^{ij}=h_i \delta^{ij}$ (the soft breaking
terms are not important in this initial inflationary stage and will be
considered later). If the couplings $f,g \sim 0.01$
\footnote{We will show later that intermediate scale breaking requires
$f\gtrsim 0.01$.}
then the condition $V(\phi)\lesssim M_{Pl}^4$ restricts the amplitudes of
${\tilde P}$ and ${\tilde P}^\prime$ to be ${\cal O}(M_{Pl})$ which is not
enough to solve the flatness and horizon problems. This leaves the
right-handed sneutrino as the only candidate for the inflaton. Clearly the
lightest sneutrino will end up being the inflaton because it is assumed to
have the flatest potential (or smallest Majorana Yukawa coupling). The
heavier generations will roll to their minimum fairly quickly because their
potential is steeper. In addition the ${\tilde P}$ and ${\tilde P}^\prime$
scalar fields receive induced masses of ${\cal O}(M_{Pl})$ and are also
driven to their minima early on. Thus if we suppose that the right-handed
electron sneutrino acts as the inflaton with ${\tilde N}_1^c(0)\gg M_{Pl}$
then during inflation the potential (\ref{vphi}) effectively becomes
\begin{equation}
	V(\phi)={1\over 4} h_1^2 \left|{\tilde N}_1^c\right|^4,
\label{infpot}
\end{equation}
with $\langle{{\tilde N}_{2,3}^c}\rangle,\langle{\tilde P}\rangle,\langle{
\tilde P}^\prime\rangle \ll M_{Pl}$. Note that the Higgs ($H_u$) and slepton
scalar fields receive induced masses from the inflaton,
$\langle{\tilde N}_1^c(t)\rangle$, which are bigger than the Hubble constant
$H$. Consequently the ${\cal O}(M_{Pl})$ amplitudes of these fields will be
damped away exponentially during the inflationary period.

The inflationary potential (\ref{infpot}) is known to generate the required
density perturbations for large scale structure of the universe, provided
that the quartic coupling ($h_1^2$) is approximately $10^{-14}$
\cite{lindebook}. This means that the Majorana Yukawa coupling for the
right-handed electron neutrino must be $h_1 \simeq 10^{-7}$. This is similar
in magnitude to the electron Yukawa coupling in the standard model, $h_e
\simeq 10^{-6}$ and suggests that the reason why the inflationary potential
is so flat is related to the (as yet) unknown reason why the electron Yukawa
coupling is very small. Given an intermediate scale breaking $\langle
{\tilde P}\rangle\simeq 10^{12}$ GeV, (see the next section) the mass scale
of the electron Majorana neutrino would be $M_{N_1}\simeq h_1\langle
{\tilde P}\rangle \simeq 10^5$ GeV. The two heavier Majorana neutrino
generations are not determined by inflation. If one assumes a hierarchy in
the Majorana Yukawa couplings similar to that of the quark and lepton mass
spectrum, an interesting light neutrino spectrum can result, with $\Delta
m_{\mu e}^2 \simeq 10^{-5} {\rm eV}^2$ and in certain cases $m_{\nu_e} >
m_{\nu_\mu}$. In principle one can also obtain an estimate for the ratio of
hot to cold dark matter.

During the initial period of chaotic inflation quantum de-Sitter
fluctuations can affect the classical motion of the inflaton. The amplitude
of the inflaton decreases exponentially during the de-Sitter phase
\cite{lindebook}
\begin{equation}
\label{iamp}
	{\tilde N}_1^c(t)={\tilde N}_1^c(0) {\rm exp}\left[-{h_1\over
	\sqrt{6\pi}} M_{Pl} t \right].
\end{equation}
After a time $\Delta t=H^{-1}$ the amplitude of the inflaton decreases by an
amount $\Delta {\tilde N}_1^c=M_{Pl}^2/(2\pi{\tilde N}_1^c)$, whereas the
average amplitude of the quantum fluctuations grows by $\left|\delta
{\tilde N}_1^c\right|=H/(2\pi)$. In order for the quantum fluctuations to
have negligible influence on the classical evolution ${\tilde N}_1^c(t)$ we
need ${\tilde N}_1^c(0)\ll h_1^{-1/3} M_{Pl}\simeq 10^2 M_{Pl}$. In addition
the universe must expand greater than 65 e-folds to solve the flatness and
horizon problems. This restricts the initial value of the inflaton field to
lie in the range $5 M_{Pl} \lesssim{\tilde N}_1^c(0)\lesssim 10^2 M_{Pl}$.

\section{PQ symmetry breaking}

The intermediate scale breaking of PQ symmetry occurs when the singlet
scalar field, ${\tilde P}$ receives a vacuum expectation value $\langle{
\tilde P}\rangle \simeq 10^{12}$ GeV. Murayama, Suzuki and Yanagida
\cite{msy} showed that this breaking can be induced by radiative corrections
from right-handed neutrino loops, which drives the mass squared parameter of
${\tilde P}$ negative. The soft supersymmetric breaking terms in the scalar
potential involving ${\tilde N}_i^c$,${\tilde P}$ and ${\tilde P}^\prime$
are given by
\begin{eqnarray}
\label{Vsoft}
	V_{soft}&=&m_{\tilde P}^2 \left|{\tilde P}\right|^2+m_{{\tilde P}^
	\prime}^2 \left|{\tilde P}^\prime\right|^2 + m_{{\tilde N}_i^c}^2
	\left|{\tilde N}_i^c\right|^2+(A_N^{(ij)} h_N^{ij}{\tilde N}_i^c
	{\tilde L}_j H_u+h.c.)\nn \\
	&+& ({\half} h_i A_i{\tilde N}_i^c {\tilde N}_i^c {\tilde P}+
	{f\over M_{Pl}}A_f{\tilde P}^3{\tilde P}^\prime+{g\over M_{Pl}}
	A_g H_u H_d {\tilde P}{\tilde P}^\prime+h.c.).
\end{eqnarray}
The soft scalar masses and trilinear couplings are all apriori unknown mass
parameters, but a study of constrained minimal supersymmetry requires them
to be ${\cal O}$(1 TeV)\cite{kkrw}. When $m_{\tilde P}^2\simeq -m_W^2$,
where $m_W\simeq {\cal O}$(TeV) is the electroweak scale, the minimum of the
scalar potential
\begin{equation}
\label{phiPpot}
	V({\tilde P})=-m_W^2 \left|{\tilde P}\right|^2 + {f^2\over M_{Pl}^2}
	           \left|{\tilde P}\right|^6 + V_0
\end{equation}
occurs for
\begin{equation}
\label{pqvev}
	\langle{\tilde P}\rangle=
	\sqrt{{m_W M_{Pl}\over \sqrt{3} f}}\simeq 10^{12} {\rm GeV}
\end{equation}
where $f\sim 0.01$ and $V_0$ is the vacuum energy associated with
the phase transition. A significantly smaller value of the coupling $f$
would increase the intermediate mass scale and conflict with cosmological
axion mass bounds \cite{twl}. In addition one finds that to stabilise the
scalar potential we need $m_{{\tilde P}^\prime}^2 > 0$ and $\langle
{\tilde P}^\prime\rangle\simeq\langle{\tilde P}\rangle$ \cite{msy}. When the
quantum corrections to the soft scalar masses in (\ref{Vsoft}) are included
via the one-loop renormalisation group equations, boundary conditions at
$M_{Planck}$ determine whether the tree-level result (\ref{pqvev}) remains
valid. In particular for $m_{\tilde P}^2$ and $m_{{\tilde N}_i^c}^2$ we have
\begin{eqnarray}
\label{rgeP}
	{d m_{\tilde P}^2 \over d t}&=&{1\over 16\pi^2}\sum_i \left|h_i
	\right|^2 (m_{\tilde P}^2+2m_{{\tilde N}_i^c}^2+\left|A_i\right|^2)
	\\
\label{rgeN}
	{d m_{{\tilde N}_i^c}^2 \over d t}&=&{1\over 16\pi^2}2\left|h_i
	\right|^2 (m_{\tilde P}^2+2m_{{\tilde N}_i^c}^2+\left|A_i\right|^2).
\end{eqnarray}
Note that in (\ref{rgeN}) we have not written slepton and Higgs soft mass
terms. A complete analysis of all the renormalisation group equations in the
MSSM which includes the neutrino masses can have interesting implications.
As we noted in the previous section $h_1\simeq {\cal O}(10^{-7})$ and
consequently its effect on the renormalisation group running (\ref{rgeP})
and (\ref{rgeN}) is negligible when $h_2,h_3 \gg h_1$. This means that the
running of $m_{{\tilde N}_2^c}^2$ and $m_{{\tilde N}_3^c}^2$ will be
identical to $m_{\tilde P}^2$. To ensure that only $m_{\tilde P}^2$ goes
negative we have to impose the boundary condition $m_{{\tilde N}_{2,3}^c}^2
\gtrsim 3 m_{\tilde P}^2$ at $M_{Planck}$. Numerical integration of the
renormalisation group equations (\ref{rgeP}) and (\ref{rgeN}) with these
boundary conditions leads to radiative PQ-symmetry breaking at an
intermediate scale, $\langle{\tilde P}\rangle\simeq 10^{12}$GeV.

The radiative corrections indicated by the renormalisation group equations
(\ref{rgeP}) and (\ref{rgeN}) are evaluated at a temperature $T=0$ and the
quantum fields are assumed to be at their minima. However we need to include
corrections arising from the inflationary period and consider possible
thermal effects. During the inflationary epoch the inflaton field sits far
from its minimum with a value ${\tilde N}_1^c(0)\gtrsim M_{Pl}$. As noted
earlier the inflaton can induce masses to any other scalar fields that it
couples to in the scalar potential (\ref{vphi}). While the Higgs and slepton
fields receive an effective mass $\gtrsim H$, the coupling $h_1^2\left|{
\tilde N}_1^c\right|^2\left|{\tilde P}\right|^2$ induces an effective mass
$h_1 \langle{\tilde N}_1^c(t)\rangle$ for ${\tilde P}$. This mass will
dominate any radiative corrections until the inflaton field
${\tilde N}_1^c(t)$ settles to its minimum after inflation ends.

The finite temperature corrections associated with the potential
(\ref{phiPpot}) have been previously discussed in the context of
intermediate scale breaking in superstring models \cite{ky}. The finite
temperature potential for $m_W \ll T\ll M_I$ and excluding the region
$T\sim\left|{\tilde P}\right|$ is given by
\begin{eqnarray}
\label{phiPtemp}
	V(\left|{\tilde P}\right|,T)&\simeq&-m_W^2\left|{\tilde P}\right|^2
	+{\pi^2\over 90}T^4 \qquad\qquad (T\ll \left|{\tilde P}\right| < M_I)
	\\
	&\simeq&{h_3^2\over 24}T^2\left|{\tilde P}\right|^2 \qquad\qquad
	\qquad\qquad (T\gg\left|{\tilde P}\right|)
\end{eqnarray}
where $M_I$ is the intermediate breaking scale. Since the third generation
Majorana neutrino Yukawa coupling $h_3 \simeq 1$, the finite temperature
potential has a local minimum at $\langle{\tilde P}\rangle =0$ which
disappears when $T\lesssim m_W \simeq 10^3$GeV. The problem is that when
the universe is supercooled at the end of inflation, the inflaton induced
${\tilde P}$ mass ($h_1 \langle{\tilde N}_1^c(t)\rangle$) still dominates
the radiative corrections ($\langle{\tilde N}_1^c\rangle\simeq{\textstyle{1
\over 3}}M_{Pl}$) and so $\langle{\tilde P}\rangle \simeq 0$. When the
universe reheats to a temperature $T_{RH}\simeq 10^4$GeV the scalar field
${\tilde P}$ is still trapped at the origin with a barrier of height
$\sim T^4$. Eventually when $T\simeq m_W$ the barrier disappears and then
$\langle{\tilde P}\rangle \simeq M_I \simeq 10^{12}$GeV. Electroweak
baryogenesis will then be the only possibility for generating a baryon
asymmetry.

However, in general there are many flat directions in supersymmetric theories
and it is very likely that a flat-direction field, $\eta$ has an amplitude
${\cal O}(M_{Pl})$. This can occur via quantum de-Sitter fluctuations along
the F and D-flat directions or it can be driven to an ${\cal O}(M_{Pl})$
local minimum by non-renormalisable Kahler potential couplings during the
initial inflationary epoch (see Dine et al \cite{drt}). If we assume this is
the case then as the right-handed electron sneutrino continues to roll
towards its minimum, there will be a point where the flat-direction field
$\eta$ dominates the potential energy density with $\eta(0)\simeq M_{Pl}$
and $V(\eta)={\half} m_W^2 \eta^2$. A second period of chaotic inflation will
then commence, which accelerates the damping of the ${\tilde N}_1^c$
oscillations and supercools the universe again. Eventually the $\tilde P$
soft term will dominate the inflaton induced ${\tilde P}$ mass (temperature
effects are negligible) and generate an intermediate scale ($\langle{
\tilde P}\rangle\simeq M_I$). We can neglect the quantum de-Sitter
fluctuations during this second period of inflation because $\sqrt{\langle
\chi^2\rangle}\lesssim m_W$. The pseudo-Nambu-Goldstone boson resulting
from the spontaneous symmetry breakdown will be the invisible axion. The
right-handed electron sneutrino and electron Majorana neutrino will then
become massive with $M_{N_1}\simeq 10^5$ GeV. In addition the MSSM Higgs
mass term $(\mu{\hat H_u}{\hat H_d})$ is generated with $\mu\simeq{\cal O}
(m_W)$.

The number of e-foldings $N$, produced during the intermediate scale
inflation depends on the initial value of the flat-direction field and is
given by $N\simeq 2\pi\eta (0)^2/M_{Pl}^2$. Typically we expect $\eta(0)
\lesssim 2 M_{Pl}$ to avoid fine-tuning problems and so $N\lesssim 25$.
This amount of inflation is not enough to expand different $\theta_i$ axion
domains beyond our present day horizon, so cosmic axion strings will be
present (although the strings will be diluted). However, even if axion
strings occur and lead to domain walls at the QCD transition temperature it
is not clear that this leads to any cosmological problems \cite{ggr}.

The adiabatic density perturbations produced by this second period of
inflation will be negligible because $\delta\rho/\rho \simeq m_W/M_{Pl}
\simeq 10^{-16}$. In order that they become irrelevant for galaxy formation
and not destroy the density perturbations produced by ${\tilde N}_1^c$, one
requires that the density perturbations re-enter the horizon for time
scales irrelevant to the growth of large scale cosmological density
perturbations. This requires that the number of e-foldings $N \leq 30$
for $T_{RH}\simeq 10^6$GeV \cite{rt} which is satisfied for $\eta(0)\simeq
2 M_{Pl}$. Note also that quantum fluctuations of the axion field can
produce isothermal density perturbations, but these will be negligible
because the Hubble constant during this second inflationary epoch is
$H\simeq m_W$ \cite{twl}.

When the second period of inflation ends the universe will be reheated
to a temperature $T_{RH}\simeq g_{\star}^{-1/4}\sqrt{\Gamma_\eta M_{Pl}}
\simeq 10^6$GeV, where $g_\star \simeq 280$ and $\Gamma_\eta\simeq h_Y^2
m_W/(4\pi)$ for an inflaton with a Yukawa-type coupling $h_Y\simeq 10^{-4}$.
Since ${\tilde P}$ is sitting (up to quantum fluctuations $\lesssim{\cal O}
(m_W)$) at the global minimum $\langle{\tilde P}\rangle \simeq M_I$ with a
potential depth $m_W^2 M_I^2\simeq (10^7 {\rm GeV})^4$, finite temperature
corrections do not destroy this minimum, even though a local minimum exists
at the origin. Note that the reheat temperature is sufficiently low to avoid
the gravitino problem \cite{gravitino}. In addition the reheat temperature
is sufficiently high that right-handed electron neutrinos ($N_1$) are
regenerated because $M_{N_1}\simeq 10^5$ GeV. When $T\simeq M_{N_1}$ a
lepton asymmetry will be generated by out of equilibrium CP-violating decays
of $N_1$, provided that the neutrino Dirac Yukawa couplings are complex and
$\left| h_N^{1j}\right|\simeq 10^{-6}$. This lepton asymmetry will be
reprocessed into a baryon asymmetry by the usual electroweak anomaly.

\section{Conclusion}

We have described how one can obtain chaotic inflation with a radiatively
generated intermediate mass scale in a simple phenomenological extension of
the MSSM. We build on but significantly modify the approach of Murayama et al
\cite{msyy}. An initial period of inflation, driven by a quartic potential
associated with the right-handed electron sneutrino solves the usual horizon
and flatness problems of the universe. Density perturbations, $\delta\rho/
\rho\simeq 10^{-5}$ are generated when the electron Majorana neutrino Yukawa
coupling is ${\cal O}(10^{-7})$. While technically natural, this coupling is
similiar in magnitude to the electron Dirac Yukawa coupling in the MSSM.
Radiative corrections from right-handed neutrino loops will break
U$(1)_{PQ}$ at an intermediate scale ($10^{12}$GeV) when the universe cools
to a temperature $T\lesssim 10^3$GeV. This implies an electron Majorana
neutrino mass $M_{N_1}\simeq 10^5$GeV and suggests that there is a hierarchy
in the Majorana neutrino spectrum related to the quark and lepton spectrum.
A baryon asymmetry can only be generated by invoking the non-perturbative
processes of electroweak baryogenesis.

However it is possible that a second stage of inflation can occur at an
intermediate scale with some flat-direction field, $\eta$. This second
inflationary epoch accelerates the damping of the electron sneutrino
amplitude and exponentially cools the universe again, causing the radiative
corrections to dominate and spontaneously break U$(1)_{PQ}$. When inflation
ends the universe can be reheated to a temperature $T_{RH}\simeq 10^6$GeV
which is sufficiently low to prevent restoring PQ symmetry. In addition the
right-handed electron neutrinos are regenerated and eventually decay when
$T\simeq M_{N_1}$. The resulting lepton asymmetry is reprocessed by the
electroweak anomaly into a baryon asymmetry.

The model we have constructed is an attempt to amalgamate current
cosmological ideas with the well tested phenomenology of the MSSM.
Ultimately one would like motivation from a more fundamental theory, but we
hope that the effective Lagrangian we have considered can shed some light
in this direction.

\section*{Acknowledgments}

We would like to thank M.~Einhorn, H.~Feldman, K.~Freese, C.~Kolda,
S.~Martin, H.~Murayama and R.~Watkins for discussions and comments. This
work was supported in part by the Department of Energy.

\end{document}